\documentclass[aos,MSNbibl,seceqn,dvips]{arximspdf}
\usepackage{mathbh}
\usepackage{accents}

\doi{10.1214/12-AOS1035} 
\volume{40}
\issue{4}
\pubyear{2012}
\firstpage{2239}
\lastpage{2265}

\makeatletter

\newproclaim{Remark}{Remark}

\newcommand{\uunderline}{\underaccent{\bar}}

\newcommand{\ddd}{\mathrm{d}}

\newtheorem{theorem}{Theorem}[section]
\newtheorem{proposition}{Proposition}[section]
\newtheorem{lemma}{Lemma}[section]
\newtheorem{corollary}{Corollary}[section]

\newcommand{\ind}[1]{\mathbh{1}_{\{#1\}}} 
\newcommand{\Expel}{\Exp_{\lambda}}
\newcommand{\Prol}{\Pro_{\lambda}}
\newcommand{\cFt}{{\mathcal{F}}_{t}}
\newcommand{\cGt}{{\mathcal{G}}_{t}}
\newcommand{\Var}{{\mathsf V}}
\newcommand{\Exp}{{\mathsf E}}
\newcommand{\Pro}{{\mathsf P}}
\newcommand{\D}{\Delta}
\newcommand{\cphi}{{\phi}}
\newcommand{\cD}{\mathcal{D}}
\newcommand{\cC}{\mathcal{C}}
\newcommand{\cF}{\mathcal{F}}
\newcommand{\cG}{\mathcal{G}}
\newcommand{\cB}{\mathcal{B}}
\newcommand{\cN}{\mathcal{N}}
\newcommand{\cM}{\mathcal{M}}
\newcommand{\cS}{\mathcal{S}}
\newcommand{\cL}{\mathcal{L}}
\newcommand{\mle}{\cS}
\newcommand{\dmle}{\tilde{\cS}}
\newcommand{\tmle}{\tilde{\lambda}}
\newcommand{\tB}{\tilde{B}}
\newcommand{\tA}{\tilde{A}}
\newcommand{\calo}{\mathcal{O}}

\makeatother

\begin{document}
\begin{frontmatter}

\title{Asymptotically optimal parameter estimation under communication constraints}
\runtitle{Parameter estimation under communication constraints}

\begin{aug}
\author[A]{\fnms{Georgios} \snm{Fellouris}\corref{}\ead[label=e1]{fellouri@usc.edu}}
\runauthor{G. Fellouris}
\affiliation{University of Southern California}
\address[A]{Department of Mathematics\\
University of Southern California \\
3620 South Vermont Ave.\\
KAP 416\\
Los Angeles, California 90089-2532 \\
USA\\
\printead{e1}} 
\end{aug}

\received{\smonth{2} \syear{2011}}
\revised{\smonth{7} \syear{2012}}

%
\begin{abstract}
A parameter estimation problem is considered, in which dispersed
sensors transmit to the statistician partial information regarding
their observations. The sensors observe the paths of continuous
semimartingales, whose drifts are linear with respect to a common
parameter. A~novel estimating scheme is suggested, according to which
each sensor transmits only one-bit messages at stopping times of its
local filtration. The proposed estimator is shown to be consistent and,
for a large class of processes, asymptotically optimal, in the sense
that its asymptotic distribution is the same as the exact distribution
of the optimal estimator that has full access to the sensor
observations. These properties are established under an asymptotically
low rate of communication between the sensors and the statistician.
Thus, despite being asymptotically efficient, the proposed estimator
requires minimal transmission activity, which is a desirable property
in many applications. Finally, the case of discrete sampling at the
sensors is studied when their underlying processes are independent
Brownian motions.
\end{abstract}

%

\end{frontmatter}

\section{Introduction}\label{sec1}
Consider a number of dispersed sensors, each one of which observes the
path of a real-valued stochastic process.
The joint distribution of these processes is assumed to belong to some
parametric family. The goal is to estimate the unknown parameter
at a central location (\textit{fusion center}) that receives
information from all sensors.

When the sensors transmit their complete observations to the fusion
center, we have a classical (\textit{centralized}) parameter
estimation problem.
However, the fusion center often does not have full access to the
sensor observations due to practical considerations, such as limited
communication bandwidth.
These communication constraints are present in applications such as
mobile and wireless communication, data fusion, environmental
monitoring and distributed surveillance, in which it is crucial to
minimize the congestion in the network and the computational burden at
the fusion center (see, e.g.,
Foresti et al.~\cite{for}).

Under this setup, which is often called \textit{decentralized}, each
sensor needs to transmit a small number of bits per communication to
the fusion center and it is clear that the classical (centralized)
statistical techniques are no longer applicable. As a result, there has
been a great interest in decentralized formulations of statistical
problems (see, e.g., the review papers by Viswanathan and Varshney
\cite{vis}, Blum et al.~\cite{blum}, Han and Amari~\cite{han} and
Veeravalli~\cite{veer}).

Parameter estimation under a decentralized setup has been studied
extensively using information-theoretic techniques. More specifically,
it is often assumed that there are two correlated sensors, each of
which observes a sequence of independent and identically distributed
(i.i.d.), finite-valued random variables whose joint probability mass
function is determined by the unknown parameter. The sensors are then
required to transmit to the fusion center messages that belong to
alphabets of smaller size than those of the original observations. The
review paper by Han and Amari~\cite{hano} describes in detail the main
advances in this line of research. On the other hand, Luo~\cite{luo}
and Xiao and Luo~\cite{luo2} considered an arbitrary number of
independent sensors that take i.i.d. observations with a common mean,
which is the unknown parameter. Assuming that the parameter space and
the support of the noise distribution are both compact intervals, they
constructed decentralized estimating schemes that require the
transmission of a small number of bits per communication.

In all the above papers, the sensors collect i.i.d. observations at a
sequence of discrete times and transmit a small number of bits to the
fusion center
at every such sampling time. Moreover, even under an asymptotically
large horizon of observations, the resulting estimators have larger
mean square errors than the corresponding optimal centralized
estimators, which have full access to the sensor observations.

In this paper the goal is to construct a decentralized estimating
scheme that requires minimal communication activity from the sensors
\textit{and} achieves asymptotically the mean square error of the
optimal centralized estimator, under a general statistical model for
the sensor observations. In particular,
we assume that the sensors observe the paths of continuous
semimartingales whose drifts are linear with respect to the unknown parameter.

The centralized version of this problem is well understood. For
Gaussian processes with independent increments, the fixed-horizon
maximum likelihood estimator (MLE) was studied by Grenander~\cite{gre}
and Striebel~\cite{str}. Brown and Hewitt~\cite{bro} proved that the
MLE is consistent and asymptotically normal
for stationary and ergodic time-homogeneous diffusions. Feigin~\cite
{fei} established the same properties for more general diffusions,
assuming that the score process is a martingale. Liptser and Shiryaev
(\cite{lip}, pages 225--236) studied the MLE for a diffusion-type
process and computed its bias and variance in the Ornstein--Uhlenbeck
case. For a diffusion-type process with linear drift with respect to
the unknown parameter, Liptser and Shiryaev~\cite{lip}, pages
244--248, and earlier Novikov~\cite{novi}, suggested a~\textit
{sequential} version of the MLE and proved that it is unbiased and that
it attains a prescribed accuracy.
In the particular case of a square root diffusion, Brown and Hewitt
\cite{bro2} suggested an alternative sequential estimator with similar
optimality properties.
Melnikov and Novikov~\cite{mel} and Galtchouk and Konev~\cite{gal}
studied least-squares sequential estimators that attain a prescribed
accuracy in a
multidimensional semimartingale regression model, generalizing in this
way the results of Novikov~\cite{novi}. We refer to Kutoyants \cite
{kuto} and Rao~\cite{rao} for exhaustive references in the statistical
inference of diffusion and diffusion-type processes.

Apart from the statistical model for the sensor observations, our work
differs from previous approaches in some other important aspects as
well. First of all, we do not assume that the frequency with which a
sensor transmits its messages to the fusion center (\textit
{communication rate}) is the same as the frequency with which it
collects its local observations (\textit{sampling rate}). Instead, we
assume that the sensors observe their underlying processes
continuously, but communicate with the fusion center at discrete times.
Therefore, in our context, the incurred loss of information is not only
due to the quantization of sensor observations, but also due to the
discrete transmission of messages to the fusion center in comparison to
the continuous flow of information at the sensors.

Moreover, we do not require that the sensors communicate with the
fusion center at deterministic and equidistant times.
Instead, we allow each sensor to transmit its messages to the fusion
center at random times that are triggered
by its local observations. In particular, we propose a communication
scheme according to which the sensors transmit only \textit{one-bit}
messages at first exit times of appropriate, locally-observed
statistics (see Rabi et al.~\cite{rabi} and Fellouris and Moustakides
\cite{fel} for similar communication schemes in different decentralized
problems). Based on this communication scheme, we construct an
estimator that is always consistent, even when the sensor processes are
dependent.

However, the main result of this paper is that, in certain cases, the
asymptotic distribution of the proposed estimator is the same as the
exact distribution of the corresponding optimal centralized estimator.
In particular, this holds when the sensor processes are arbitrary,
orthogonal continuous semimartingales, as well as when they are
correlated Gaussian processes with independent increments.

More importantly, these asymptotic properties are established as the
horizon of observations goes to infinity and as the rate of\vadjust{\goodbreak}
communication between sensors and the fusion center goes to
\textit{zero}.
Thus, although the proposed estimator is statistically efficient,
it requires minimal communication activity from the sensors, which is a
very desirable property in applications with severe communication constraints.

Finally, we consider in more detail the special case in which the
sensors observe independent Brownian motions, since the tractability of
this model allows us to obtain additional insight regarding the
suggested estimating scheme. In this context, we also consider the case
of discrete sampling, where the sensors do not observe their underlying
processes continuously, but at a sequence of discrete times. It is
shown that the proposed estimator remains consistent for any fixed
sampling frequency, as long as the sensors have an asymptotically low
rate of communication with the fusion center. However, asymptotic
optimality does require
a sufficiently high sampling rate, which we determine as a function of
the communication rate and the observation horizon.

The rest of the paper is organized as follows: in Section~\ref{sec2} we
formulate the problem under consideration. In Section~\ref{sec3} we
specify the
proposed estimating scheme and analyze its asymptotic properties. In
Section~\ref{sec4} we focus on the special case that the sensors observe
independent Brownian motions. We conclude in Section~\ref{sec5}.

\section{Problem formulation}\label{sec2}
In what follows, we denote by $i$ the generic sensor, where
$i=1,\ldots, K$. We assume that sensor $i$ observes the path of a
continuous stochastic process $Y^{i}=\{Y_{t}^{i}\}_{t \geq0}$ and
is\vspace*{1pt}
able to compute any statistic that is adapted to the filtration
generated by $Y^{i}$.

In this section we specify the dynamics of $(Y^{1},\ldots, Y^{K})$
under a family of probability measures
$\{\Prol, \lambda\in\mathbb{R}\}$, we review standard results
regarding the centralized estimation of the unknown parameter $\lambda
$ and
we define the notion of an (asymptotically optimal) decentralized estimator.

\subsection{Statistical model}\label{sec2.1}
Let $(Y^{1},\ldots, Y^{K})$ be the coordinate process on the canonical
space of continuous functions $(\Omega, \cF)$, where $\Omega
:=\mathbb
{C}[0,\infty)^{K}$ and $\cF:=\cB(\Omega)$ is the associated Borel
$\sigma$-algebra. We denote by $\{\cFt^{i}\}$ the right-continuous
version of
the natural filtration generated by $Y^{i}$ and by $\{\cFt\}$ the
corresponding global filtration
%
%
\begin{eqnarray}
\label{fili}
\cFt^{i} &:=& \cC^{i}_{t+},\qquad \cC_{t}^{i}:=
\sigma\bigl(Y_{s}^{i}; 0 \leq s \leq t\bigr),
\\
\label{fil}
\cFt&:=& \cC_{t+},\qquad \cC_{t}:=\sigma\bigl(Y_{s}^{i}; 0 \leq s \leq t,
1\leq i \leq K\bigr).
\end{eqnarray}
Let also $\Pro_{0}$ be a probability measure on $(\Omega, \cF)$ so that
\[
Y^{i} \in\cM_{0} \qquad\forall1 \leq i \leq K,
\]
where $\cM_{0}$ is the class of continuous $\Pro_{0}$-local
martingales that start from 0.

For every $1\leq i, j \leq K$, we denote by $\langle Y^{i}, Y^{j}
\rangle$ the quadratic covariation of $Y^{i}$ and $Y^{j}$
and we assume that $X^{i}$ is an $\{\cFt^{i}\}$-progressively
measurable process so that
%
%
\begin{equation}
\Pro_{0} \Biggl( \sum_{i=1}^{K}
\int_{0}^{t} \bigl|X_{s}^{i}\bigr|^{2}
\,\ddd \bigl\langle Y^{i}, Y^{i} \bigr\rangle_{s} <
\infty\Biggr)=1 \qquad\forall0 \leq t < \infty.
\end{equation}
Then, we can define the stochastic integral
%
%
\begin{equation}
\label{B} B_{t}:= \sum_{i=1}^{K}
\int_{0}^{t} X_{s}^{i}
\,\ddd Y_{s}^{i},\qquad t \geq0,
\end{equation}
and we denote by $A$ its quadratic variation, that is,
%
%
\begin{equation}
\label{A} A_{t}:= \langle B, B \rangle_{t} = \sum
_{i=1}^{K} \sum_{j=1}^{K}
\int_{0}^{t} X_{s}^{i}
X_{s}^{j} \,\ddd \bigl\langle Y^{i}, Y^{j}
\bigr\rangle_{s},\qquad t \geq0.
\end{equation}
Moreover, we assume that the Novikov-type condition:
{\renewcommand{\theequation}{A1}
\begin{equation}\label{equA1}
\Exp_{0} \bigl[ e^{(\lambda^2/2)
A_{t}}\bigr] < \infty\qquad\forall0 \leq t < \infty
\end{equation}}

\noindent is satisfied for every $\lambda\neq0$, which allows us to define for
every $\lambda\neq0$ the probability measure $\Prol$ in the
following way:
%
%
\setcounter{equation}{5}
\begin{equation}
\label{likeli} \frac{\mathrm{d} \Prol}{\mathrm{d} \Pro_{0}} \bigg|_{\cFt}:=
e^{\lambda
B_{t} - (\lambda^{2}/2) A_{t}} \qquad\forall0
\leq t < \infty.
\end{equation}
Then, if we denote by $\cM_{\lambda}$ the class of continuous $\Prol
$-local martingales that start from 0, Girsanov's theorem (see \cite
{rev}, page 331)
implies that
%
%
\begin{equation}
\label{N} N^{i}:= Y^{i} - \bigl\langle Y^{i},
\lambda B \bigr\rangle\in\cM_{\lambda} \qquad\forall i=1,\ldots, K
\end{equation}
and, consequently, $\langle N^{i}, N^{j} \rangle= \langle Y^{i}, Y^{j}
\rangle$ for every $i \neq j$. Therefore, from (\ref{B}) and (\ref{N})
it follows that under $\Prol$
%
%
\begin{equation}
\label{model} Y_{t}^{i} = \lambda\sum
_{j=1}^{K} \int_{0}^{t}
X_{s}^{j} \,\ddd\bigl\langle Y^{i}, Y^{j}
\bigr\rangle_{s} +N_{t}^{i},\qquad t \geq0, 1 \leq i
\leq K.
\end{equation}

\subsection{The parameter estimation problem}\label{sec2.2}
The goal is to estimate the unknown parameter $\lambda$ using the
information that is being transmitted from the sensors to the fusion center.
The flow of this information can be described by a sub-filtration of
$\{\cFt\}$ and is determined by the \textit{communication scheme}
that is chosen by the statistician.

Let $\{\cGt\} \subset\{\cFt\}$ be the fusion center filtration. We
will say that:
\begin{longlist}[(a)]
\item[(a)] $(\phi_{t})_{t > 0}$ is a \textit{fixed-horizon}, $\{\cGt\}
$-adapted estimator of $\lambda$, if
$\phi_{t}$ is a $\cG_{t}$-measurable statistic for every $t>0$.\vadjust{\goodbreak}
\item[(b)] $(T_{\gamma}, \phi_{\gamma})_{\gamma>0}$ is a \textit
{sequential}, $\{\cGt\}$-adapted estimator of $\lambda$, if
$(T_{\gamma})_{\gamma>0}$ is an increasing family of $\{\cGt\}
$-stopping times and $\cphi_{\gamma}$ a $\cG_{T_{\gamma}}$-measurable
statistic
for every $\gamma> 0$.
\end{longlist}
We will say that a $\{\cGt\}$-adapted estimator, either fixed-horizon
or sequential, is \textit{decentralized}, when the fusion center
filtration $\{\cGt\}$
is of the form
%
%
\begin{equation}
\label{ff} \cGt=\sigma\bigl(\sigma_{n}^{i},
\chi_{n}^{i}| \sigma_{n}^{i} \leq t,
i=1,\ldots,K\bigr),\qquad t \geq0,
\end{equation}
where $(\sigma_{n}^{i})_{n \in\mathbb{N}}$ is an increasing sequence
of $\{\cFt^{i}\}$-stopping times and each
$\chi_{n}^{i}$ is an $\cF^{i}_{\sigma_{n}^{i}}$-measurable\vspace*{1pt} statistic
that takes values in a \textit{finite} set.
In other words, a decentralized estimator must rely on quantized
versions of the sensor observations, which may be transmitted to the
fusion center
at stopping times of the local sensor filtrations.

If the fusion center learns the complete sensor observations at any
time $t$, then it can construct $\{\cFt\}$-adapted estimators, which
we will call \textit{centralized}. Assuming that for every $\lambda
\in
\mathbb{R}$,
{\renewcommand{\theequation}{A2}
\begin{equation}\label{equA2}
\Prol(A_{t}>0)=1\qquad\forall t>0,
\end{equation}}

\noindent the centralized, fixed-horizon MLE of $\lambda$ at some time $t>0$ is
%
%
\setcounter{equation}{9}
\begin{equation}
\label{mle} \hat{\lambda}_{t}:= \frac{B_{t}}{A_{t}},
\end{equation}
that is, the maximizer of the corresponding log-likelihood function,
%
%
\begin{equation}
\label{loglikeli} \ell_{t}(\lambda):= \log\frac{\mathrm{d}\Prol}{\mathrm
{d}\Pro_{0}}
\bigg|_{\cF_{t}} = \lambda B_{t} - \frac{\lambda^{2}}{2} A_{t}.
\end{equation}
From (\ref{loglikeli}) we also obtain the corresponding score process
and (observed) Fisher information, that is,
%
%
\begin{equation}
\label{score0} M_{t}:= \frac{\mathrm{d} \ell_{t}(\lambda)}{\mathrm{d}
\lambda
} = B_{t} - \lambda
A_{t},\qquad -\frac{\mathrm{d}^{2} \ell_{t}(\lambda)}{\mathrm{d} \lambda^{2}}
= A_{t},\qquad t \geq0,
\end{equation}
and, consequently, we have
%
%
\begin{equation}
\label{deco2} \hat{\lambda}_{t}= \lambda+ \frac{M_{t}}{A_{t}},\qquad t > 0.
\end{equation}
Moreover, from (\ref{B}), (\ref{A}) and (\ref{model}) it follows that
$M\in\cM_{\lambda}$, since
%
%
\begin{equation}
\label{score1} M_{t}= \sum_{i=1}^{K}
\int_{0}^{t} X_{s}^{i}
\,\ddd N^{i}_{s}, \qquad t \geq0.
\end{equation}
Since $\langle M, M \rangle= \langle B, B \rangle=A$, if we also assume
that for every $\lambda\in\mathbb{R}$
{\renewcommand{\theequation}{A3}
\begin{equation}\label{equA3}
\Prol\Bigl(\lim_{t \rightarrow\infty} A_{t}=\infty\Bigr)=1,
\end{equation}}
then there exists a $\Prol$-Brownian motion $W$ (see~\cite{kar}, page
174) so that
%
%
\setcounter{equation}{14}
\begin{equation}
\label{dds} \Prol(M_{t} =W_{A_{t}}, t \geq0)=1.
\end{equation}
This representation has some important consequences, which we state in
the following lemma.
%
%
\begin{lemma} \label{propo1}
\textup{(a)} If $(t_{\gamma})_{\gamma>0}$ is an increasing family of
(possibly random) times so that
$t_{\gamma} \rightarrow\infty$ $\Prol$-a.s., then $\hat{\lambda
}_{t_{\gamma}} \rightarrow\lambda$ $\Pro_{\lambda}$-a.s. as
$\gamma
\rightarrow\infty$.

\textup{(b)} If $T_{1} \leq T_{2}$ are $\{\cF_{t}\}$-stopping times so
that $\Expel[A_{T_{2}}]< \infty$, then
%
%
\begin{eqnarray}
\label{wald1}
\Expel[M_{T_{1}}] &=& \Expel[M_{T_{2}}]=0,
\\
\label{wald2}
\Expel\bigl[(M_{T_{2}}-M_{T_{1}})^{2}\bigr]&=&
\Expel[A_{T_{2}}-A_{T_{1}}].
\end{eqnarray}

\textup{(c)} If $\{A_{t}\}$ is deterministic, then
%
%
\begin{equation}
\label{asydet} \sqrt{A_{t}} (\hat{\lambda}_{t}-\lambda) \sim
\cN(0,1) \qquad\forall t > 0.
\end{equation}
\end{lemma}
\begin{pf}
Part (a) is a consequence of (\ref{deco2}), (\ref{dds}) and the strong
law of large numbers for the Brownian motion. Part (b) follows from a
localization argument, optional sampling theorem and Doob's maximal
inequality. Finally, when $\{A_{t}\}$ is deterministic, from (\ref
{dds}) it follows that $M_{t} \sim\cN(0,A_{t})$ for every $t >0$. From
this observation and (\ref{deco2}) we obtain (\ref{asydet}).
\end{pf}

In the following lemma we state a version of the Cramer--Rao--Wolfowitz
inequality.
%
%
\begin{lemma} \label{crw}
If $T$ is an $\{\cFt\}$-stopping time and $\phi$ is an $\cF
_{T}$-measurable statistic so that
$0 < \Expel[A_{T}]<\infty$ and $\Expel[\phi]=\lambda, \Var_{\lambda
}[\phi]<\infty$ for every $\lambda\in\mathbb{R}$, then
\[
\Var_{\lambda}[\phi] \geq\frac{1}{\Expel[A_{T}]}.
\]
\end{lemma}
\begin{pf}
From (\ref{wald1}) and (\ref{wald2}) and the Cauchy--Schwarz inequality
we have
\[
\Expel[ \phi M_{T}] = \Expel\bigl[(\phi-\lambda) M_{T}
\bigr] \leq\sqrt{ \Expel\bigl[(\phi-\lambda)^{2}\bigr] \Expel
\bigl[(M_{T})^{2} \bigr]} = \sqrt{ \Var_{\lambda}[\phi]
\Expel[A_{T}]}.
\]
Thus, it suffices to show that $\Expel[\phi M_{T}] =1$. Indeed,
changing the measure $\Prol\mapsto\Pro_{0}$ and differentiating both
sides in $\Expel[\phi]=\lambda$ with respect to $\lambda$,
\[
1 = \frac{\mathrm{d}}{\mathrm{d} \lambda} \Exp_{0} \bigl[ e^{\lambda
B_{T}- (\lambda^{2}/2) A_{T}} \phi\bigr] =
\Exp_{0} \bigl[ e^{\lambda B_{T}- (\lambda^{2}/2) A_{T} } M_{T} \phi
\bigr] = \Expel[
M_{T} \phi].
\]
The second equality follows from interchanging derivative and
expectation, which is possible due to the (quadratic) form of the
log-likelihood function (\ref{loglikeli}) (see, e.g., \cite
{kuto}, page 54).
\end{pf}

Lemma~\ref{crw} and (\ref{asydet}) imply that when $A_{t}$ is
deterministic, $\hat{\lambda}_{t}$ is an optimal estimator of
$\lambda
$, in the sense that\vadjust{\goodbreak}
it has the smallest possible variance among $\cFt$-measurable,
unbiased estimators (for any fixed $t >0$). In order to obtain such an
exact optimality property when
$\{A_{t}\}$ is random, we consider the following sequential version of
the centralized MLE:
%
%
\begin{equation}
\label{smle} \cS_{\gamma}:= \inf\{ t \geq0\dvtx A_{t}
\geq\gamma\},\qquad
\hat
{\lambda}_{\cS_{\gamma}} = \biggl( \frac{B}{A}
\biggr)_{\cS_{\gamma}},\qquad
\gamma>0.
\end{equation}
%

\begin{lemma} \label{coll}
For every $\gamma>0$,
%
%
\begin{eqnarray}
\label{no}
\Prol(\cS_{\gamma} <\infty)&=&1,
\\
\label{asyran}
\sqrt{\gamma}(\hat{\lambda}_{\cS_{\gamma}}-\lambda) &\sim& \cN(0,1).
\end{eqnarray}
Moreover, $\Prol(\hat{\lambda}_{\cS_{\gamma}} \rightarrow\lambda)=1$
as $\gamma\rightarrow\infty$.
\end{lemma}
\begin{pf}
Assumption (\ref{equA3}) implies (\ref{no}). Since $A$ has continuous paths,
$A_{\cS_{\gamma}}= \gamma$. Thus, from (\ref{dds}) we have
$M_{\cS_{\gamma}} \sim\cN(0, \gamma)$ and, consequently, from
(\ref
{asydet}) we obtain (\ref{asyran}). Finally, the strong consistency of
$\hat{\lambda}_{\cS_{\gamma}}$ as $\gamma\rightarrow\infty$ is
implied by Lemma~\ref{propo1}(a).
\end{pf}

From Lemmas~\ref{crw} and~\ref{coll} it follows that, for any given
$\gamma>0$, $\hat{\lambda}_{\cS_{\gamma}}$ is an optimal estimator of
$\lambda$,
in the sense that it has the smallest possible variance among unbiased,
$\{\cFt\}$-adapted estimators $(T_{\gamma}, \phi_{\gamma})$ for which
$\Expel[A_{T_{\gamma}}] \leq\gamma$.\vspace*{1pt}

Therefore, there is always a centralized estimator of $\lambda$ that is
unbiased,
normally distributed and optimal in a \textit{nonasymptotic} sense. A
decentralized estimator cannot have such a strong optimality property,
as it relies on less information. However, we will say that a
(decentralized) estimator is \textit{asymptotically optimal},
if it has the same distribution as the corresponding optimal
centralized estimator when an asymptotically large horizon of
observations is available. More specifically,
\begin{longlist}[(a)]
\item[(a)] when $\{A_{t}\}$ is deterministic, a fixed-horizon, $\{
\cGt
\}$-adapted estimator $(\phi_{t})_{t > 0}$ will be \textit
{asymptotically optimal} if
\[
\sqrt{A_{t}} (\phi_{t}- \lambda) \rightarrow\cN(0,1)\qquad
\mbox{as } t \rightarrow\infty,
\]

\item[(b)] when $\{A_{t}\}$ is random, a sequential, $\{\cGt\}
$-adapted estimator $(T_{\gamma}, \phi_{\gamma})_{\gamma>0}$ will be
\textit{asymptotically optimal} if
\[
\limsup_{\gamma\rightarrow\infty} \bigl(\Expel[A_{T_{\gamma}}] -\gamma
\bigr) \leq0
\quad\mbox{and}\quad \sqrt{\gamma} (\phi_{\gamma}- \lambda) \rightarrow\cN(0,1)
\qquad\mbox{as } \gamma\rightarrow\infty.
\]
\end{longlist}

\subsection{Notation}\label{sec2.3} We close this section with some
notation that
will be useful in the construction and analysis of the proposed
estimating scheme.
Thus, for every $1 \leq i \leq K$ we define the statistic
%
%
\begin{equation}
\label{Bi} B^{i}_{t}:= \int_{0}^{t}
X_{s}^{i} \,\ddd Y_{s}^{i},\qquad t \geq0,
\end{equation}
and for any $1\leq i,j \leq K$ we denote by $A^{ij}$ the quadratic
covariation of $B^{i}$ and $B^{j}$ and by $A^{i}$ the quadratic
variation of $B^{i}$,
that is,
%
%
\begin{eqnarray}
\label{Aij}
A^{ij}_{t} &:=& \bigl\langle B^{i}, B^{j}
\bigr\rangle_{t}= \int_{0}^{t}
X_{s}^{i} X_{s}^{j} \,\ddd\bigl\langle
Y^{i}, Y^{j} \bigr\rangle_{s},\qquad t \geq0,
\\[-2pt]
\label{Ai}
A^{i}_{t} &:=& \bigl\langle B^{i}, B^{i}
\bigr\rangle_{t}= \int_{0}^{t}
\bigl(X_{s}^{i}\bigr)^{2} \,\ddd\bigl\langle
Y^{i}, Y^{i} \bigr\rangle_{s},\qquad t \geq0.
\end{eqnarray}
Then, recalling the definitions of $B$ and $A$ in (\ref{B}) and (\ref{A}),
we have
%
%
\begin{equation}
\label{iso} B=\sum_{i=1}^{K} B^{i},\qquad A=\sum_{i=1}^{K} A^{i}
+\sum_{1 \leq i \neq j \leq K} A^{ij}.
\end{equation}
Moreover, we define the set
%
%
\begin{equation}
\label{set} \cD:=\bigl\{ (i,j)| 1 \leq i \neq j \leq K \mbox{ and }
A^{ij} \mbox{ is random} \bigr\}
\end{equation}
and we have the following representation for $A$:
%
%
\begin{equation}
\label{is} A=\sum_{i=1}^{K} A^{i}
+\sum_{(i,j) \in\cD} A^{ij} +\sum
_{(i,j)
\notin
\cD} A^{ij}.
\end{equation}

\section{A decentralized estimating scheme}\label{sec3}
In this section we construct and analyze the proposed decentralized
estimator. More specifically, we first define the communication scheme
at the sensors and then introduce the statistics and estimators that
will be used by the fusion center. As in the centralized setup, we
distinguish two cases and consider a fixed-horizon estimator when $\{
A_{t}\}$ is deterministic and a sequential estimator when $\{A_{t}\}$
is random. In each case, we analyze the asymptotic behavior of the
resulting estimator as the horizon of observations goes to infinity
\textit{and} the rate of communication goes to zero, assuming that conditions
(\ref{equA1}), (\ref{equA2}), (\ref{equA3}) are satisfied.

The main idea in the suggested communication scheme is that each sensor
should inform the fusion center about the sufficient statistics for
$\lambda$ that it observes locally. However, instead of communicating
at deterministic times, its communication times should be triggered by
its local observations. In other words, each sensor $i$ should inform
the fusion center about the evolution of the $\{\cFt^{i}\}$-adapted,
sufficient statistics for $\lambda$ at a sequence of $\{\cFt^{i}\}
$-stopping times.

When $A$ is deterministic, $B^{1},\ldots, B^{K}$ are the only
sufficient statistics for $\lambda$ and it is clear that each $B^{i}$
is $\{\cFt^{i}\}$-adapted, thus observable at sensor $i$.

When $A$ is random, there are additional sufficient statistics, the
\textit{random} processes of the form $A^{i}$ or $A^{ij}$ (when $A^{i}$
or $A^{ij}$ is deterministic, it is completely known to the fusion
center at any time $t$). If $A^{i}$ is random, it is clear that it is
$\{\cFt^{i}\}$-adapted, since $\langle B^{i}, B^{i} \rangle= A^{i}$.
On the other hand, if $A^{ij}$ (with $i \neq j$) is random, it is not\vadjust{\goodbreak}
locally observed either at sensor $i$ or at sensor $j$, thus, the
fusion center cannot be informed about its evolution (since there is no
communication between sensors).

\subsection{Communication scheme and fusion center statistics}\label{sec3.1}
Based on the previous discussion, we suggest that each sensor $i$
communicate with the fusion center at the times
%
%
\begin{equation}
\label{tauB} \tau_{n}^{i,B}:= \inf\bigl\{ t \geq
\tau_{n-1}^{i,B}\dvtx B_{t}^{i}-
B^{i}_{\tau_{n-1}^{i,B}} \notin\bigl( -\underline{\Delta}^{i},
\overline{\Delta}{}^{i}\bigr) \bigr\},\qquad n \in\mathbb{N},
\end{equation}
and, \textit{if $A$ and $A^{i}$ are random}, also at the times
%
%
\begin{equation}
\label{tauA} \tau_{n}^{i,A}:= \inf\bigl\{ t \geq
\tau_{n-1}^{i,A}\dvtx A^{i}_{t} -
A^{i}_{\tau_{n-1}^{i,A}} \geq c^{i} \bigr\},\qquad n \in\mathbb{N},
\end{equation}
where $\tau_{0}^{i,A}=\tau_{0}^{i,B}:=0$ and $c^{i},\overline{\Delta
}{}^{i}, \underline{\Delta}^{i}>0$ are
arbitrary, constant thresholds, chosen by the designer of the scheme,
known both at sensor $i$ and the fusion center. If either $A^{i}$ or
$A$ is deterministic, sensor $i$ does not communicate at the times
$(\tau_{n}^{i,A})$
and we set $\tau_{n}^{i,A}=\infty$ for every $n\geq1$.

At $\tau_{n}^{i,B}$, sensor $i$ transmits to the fusion center with one
bit the outcome of the Bernoulli random variable
%
%
\begin{equation}
\label{zB} z_n^i:=\cases{1, &\quad if $B^{i}_{\tau_{n}^{i,B}}-
B^{i}_{\tau
_{n-1}^{i,B}} \geq\overline{\Delta}{}^{i}$,
\vspace*{2pt}\cr
0, &\quad if
$B^{i}_{\tau_{n}^{i,B}}- B^{i}_{\tau_{n-1}^{i,B}} \leq-\underline{
\Delta}{}^{i}$,}
\end{equation}
whereas at $\tau_{n}^{i,A}$, if needed, it informs the fusion center
with one bit that $A^{i}$ has increased by $c^{i}$ since $\tau
_{n-1}^{i,A}$. Therefore, the induced filtration at the fusion center is
%
%
\begin{equation}
\label{tfil} \tilde{\cFt}:=\sigma\bigl( \tau_{n}^{i,A},
\tau_{n}^{i,B}, z_{n}^{i} |
\tau_{n}^{i,A} \leq t, \tau_{n}^{i,B} \leq
t, i=1,\ldots, K \bigr),\qquad t \geq0,
\end{equation}
which means that the fusion center can compute any $\{\tilde{\cFt}\}
$-adapted statistic.

For every $1 \leq i \leq K$ we define
%
%
\begin{eqnarray}
\label{tAi}
\tA_{t}^{i} &:=& n c^{i},\qquad \tau_{n}^{i,A}
\leq t < \tau_{n+1}^{i,A},\qquad n \in\mathbb{N} \cup\{0\},
\\
\label{tBi}
\tB^{i}_{t} &:=& \sum_{j=1}^{n}
\bigl[\overline{\Delta}{}^{i}z_{j}^{i} -\underline{
\Delta}^{i}\bigl(1-z_{j}^{i}\bigr)\bigr],\qquad
\tau_{n}^{i,B} \leq t < \tau_{n+1}^{i,B},\qquad n
\geq1,
\end{eqnarray}
where $\tB^{i}_{t}:=0$ for $t<\tau_{1}^{i,B}$, with the understanding
that $\tA^{i}:=A^{i}$ when $A^{i}$ is deterministic. Moreover,
motivated by (\ref{iso})--(\ref{is}), we define
%
%
\begin{eqnarray}
\label{tB}
\tB:\!&=& \sum_{i=1}^{K} \tilde{B}^{i},
\\
\label{tA}
\tA:\!&=& \sum_{i=1}^{K} \tA^{i} +
\sum_{i=1}^{K} d_{i}
\tA^{i} + \sum_{(i,j) \notin\cD} A^{ij}
\nonumber\\[-8pt]\\[-8pt]
&=& \sum_{i=1}^{K} (1+d_{i})
\tA^{i} + \sum_{(i,j) \notin\cD} A^{ij},
\nonumber
\end{eqnarray}
where $d_{i}$ is the number of random terms of the form $A^{ij}$, that is,
%
%
\begin{equation}
\label{di} d_{i}:= \#\bigl\{j | 1 \leq i \neq j \leq K \mbox{ and }
A^{ij} \mbox{ is random} \bigr\}.
\end{equation}
Again, we set $\tA:=A$ when $A$ is deterministic. Finally, we define
the following quantities:
%
%
\begin{equation}
\label{Dc} \D:=\sum_{i=1}^{K} \max\bigl\{
\overline{\Delta}{}^{i}, \underline{\Delta}^{i}\bigr\},\qquad c:=
\sum_{i=1}^{K} (1+d_{i})
c^{i},
\end{equation}
which will play an important role in the asymptotic analysis of the
proposed estimating scheme.
%
%
\begin{lemma} \label{lemma0}
For every $1 \leq i \leq K$ and $t,c, \D> 0$,
%
%
\begin{eqnarray}
\label{dist}
&\displaystyle 0 \leq A^{i}_{t} - \tA^{i}_{t} \leq
c^{i},\qquad \bigl|B^{i}_{t}-\tB^{i}_{t}\bigr|
\leq\max\bigl\{\overline{\Delta}{}^{i},\underline{\Delta}^{i}
\bigr\},&
\\
\label{dist2}
&\displaystyle A_{t} - \tA_{t} \leq c,\qquad |B_{t}-
\tB_{t}| \leq\D.&
\end{eqnarray}
\end{lemma}
\begin{pf}
If $A^{i}$, $A$ are deterministic, then $\tA^{i}:=A^{i}$, $\tA:=A$ and
the corresponding inequalities\vspace*{1pt} hold trivially.
Thus, without loss of generality, we assume that both $A^{i}$ and $A$
are random.

First\vspace*{1pt} of all, we observe that $\tB^{i}$ is exactly equal to $B^{i}$ at
$\tau_{n}^{i,B}$ and
$\tA^{i}$ is exactly equal to $A^{i}$ at $\tau_{n}^{i,A}$ for every
$n\in\mathbb{N}$. Indeed, due to the path continuity of $A^{i}$ and~$B^{i}$,
for every $n\in\mathbb{N}$ it is
\begin{eqnarray*}
\tA^{i}_{\tau_{n}^{i,A}} &=& n c^{i}= \sum
_{j=1}^{n}\bigl[A^{i}_{\tau
_{j}^{i,A}}-
A^{i}_{\tau_{j-1}^{i,A}}\bigr] = A^{i}_{\tau_{n}^{i,A}},
\\
\tB^{i}_{\tau_{n}^{i,B}} &=& \sum_{j=1}^{n}
\bigl[\overline{\Delta}{}^{i}z_{j}^{i}- \underline{
\Delta}^{i} \bigl(1-z_{j}^{i}\bigr)\bigr]= \sum
_{j=1}^{n}\bigl[B^{i}_{\tau_{j}^{i,B}}-
B^{i}_{\tau
_{j-1}^{i,B}}\bigr]= B^{i}_{\tau_{n}^{i,B}}.
\end{eqnarray*}
Moreover, from the definition of the communication times $(\tau
_{n}^{i,B})_{n}$, it is clear that $|B_{t}^{i}- \tB_{t}^{i}| < \max\{
\overline{\Delta}{}^{i},\underline{\Delta}^{i}\}$ for any time $t$
between two jump times of $\tB^{i}$,
which proves the second inequality in (\ref{dist}). Similarly, from the
definition of $(\tau_{n}^{i,A})_{n}$ and the fact that $A^{i}$ has
increasing paths, it is clear that $0< A_{t}^{i}- \tA_{t}^{i} < c^{i}$
for any time $t$ between two jump times of $\tA^{i}$, which proves the
first inequality in (\ref{dist}).

The second inequality in (\ref{dist2}) follows directly from the second
inequality in (\ref{dist}) and the definition of $\D$. Finally, from
the Kunita--Watanabe inequality (see~\cite{kar}, page 142) and the
algebraic inequality $2\sqrt{|xy|} \leq|x| + \break |y|$ we have
\[
\bigl|A^{ij}\bigr| \leq\sqrt{A^{i} A^{j}} \leq
\tfrac{1}{2} \bigl(A^{i} + A^{j}\bigr),\qquad 1 \leq i \neq j
\leq K,
\]
thus, from the definitions of $\cD$ and $d_{i}$ [recall (\ref{set}) and
(\ref{di})] we obtain
\[
\sum_{(i,j) \in\cD} A^{ij} \leq\frac{1}{2}
\sum_{(i,j) \in\cD} \bigl(A^{i} + A^{j}
\bigr) = \sum_{(i,j) \in\cD} A^{i}= \sum
_{i=1}^{K} d_{i} A^{i}.
\]
From the representation of $A$ in (\ref{is}) and the latter inequality
we have
\begin{eqnarray*}
A &\leq& \sum_{i=1}^{K} A^{i} +
\sum_{i=1}^{K} d_{i}
A^{i} + \sum_{(i,j) \notin\cD} A^{ij}
\\
&\leq& \sum_{i=1}^{K} (1+ d_{i})
\bigl(\tA^{i}+c^{i}\bigr) + \sum_{(i,j)
\notin\cD}
A^{ij} = \tA+ c,
\end{eqnarray*}
where the second inequality is due to (\ref{dist}) and the equality
follows from the definitions of $\tA$ and $c$ in (\ref{tA}) and
(\ref{Dc}), respectively.
\end{pf}

\subsection{The proposed estimator}\label{sec3.2}
The proposed communication scheme requires the transmission of only one
bit whenever a sensor communicates with the fusion center.
Thus, the overall communication activity in the network will be low as
long as the communication rate of each sensor is low.
Therefore, we should ideally design an $\{\tilde{\cFt}\}$-adapted
estimator that is statistically efficient even under an
asymptotically low communication rate as the horizon of observations
goes to infinity. For this reason, we let $\D\rightarrow\infty$ and
$c \rightarrow\infty$ as $t \rightarrow\infty$ (or $\gamma
\rightarrow\infty$) and we determine the relative rates that guarantee
consistency and asymptotic optimality.

When $\{A_{t}\}$ is \textit{deterministic}, we suggest the following
estimator of $\lambda$ at some arbitrary, deterministic time $t>0$:
%
%
\begin{equation}
\label{dbmled} \tilde{\lambda}_{t}:= \frac{\tB_{t}}{A_{t}}.
\end{equation}
In the following theorem, which is the first main result of this paper,
we show that $\{\tilde{\lambda}_{t}\}$ is consistent and asymptotically
optimal under an asymptotically low communication rate.
%
%
\begin{theorem} \label{theo0}
If $t, \D\rightarrow\infty$ so that $\D= o(A_{t})$, then $\tilde
{\lambda}_{t}$ converges to $\lambda$ almost surely and in mean square.
If additionally $\D= o(\sqrt{A_{t}})$, then $\tilde{\lambda}_{t}$ is
asymptotically optimal, that is, $\sqrt{A_{t}} ( \tilde{\lambda}_{t} -
\lambda) \rightarrow\cN(0,1)$.
\end{theorem}
\begin{pf}
Since $\hat{\lambda}_{t}$ converges to $\lambda$ almost surely and in
mean square as $t \rightarrow\infty$,
in order to prove that $\tilde{\lambda}_{t}$ is consistent, it suffices
to show that $\Prol(|\tilde{\lambda}_{t}-\hat{\lambda}_{t}|
\rightarrow
0)=1$ and $\Expel[ (\tilde{\lambda}_{t}-\hat{\lambda}_{t})^{2}]
\rightarrow0$ as $t, \Delta\rightarrow\infty$ so that $\D= o(A_{t})$.\vadjust{\goodbreak}

Moreover, since $\sqrt{A_{t}} (\hat{\lambda}_{t}-\lambda) \sim
\cN
(0,1)$ for any $t>0$, in order to establish the asymptotic optimality of
$\tilde{\lambda}_{t}$, it suffices to show that $\sqrt{A_{t}} |
\tilde{\lambda}_{t}-\hat{\lambda}_{t}|$ converges to 0 in probability
as $t, \D\rightarrow\infty$ so that $\D=o(\sqrt{A_{t}})$.

Indeed, from the second inequality in (\ref{dist2}) we have
\[
|\tilde{\lambda}_{t}-\hat{\lambda}_{t}|= \biggl|
\frac{\tB_{t}}{A_{t}}- \frac{B_{t}}{A_{t}} \biggr| = \frac{|\tB_{t} -
B_{t}|}{A_{t}} \leq
\frac{\D}{A_{t}},\qquad t> 0,
\]
which proves both claims.
\end{pf}

When $\{A_{t}\}$ is \textit{random}, we suggest the following
sequential, $\{\tilde{\cFt}\}$-adapted estimator of $\lambda$:
%
%
\begin{equation}
\label{dsmle} \tilde{\cS}_{\gamma}:= \inf\{ t \geq0\dvtx \tA_{t}
\geq\gamma- c \}, \qquad\tmle_{\dmle_{\gamma}}:= \biggl( \frac{\tB}{\tA}
\biggr)_{\dmle
_{\gamma
}}, \qquad\gamma>c.
\end{equation}

%
\begin{lemma} \label{leme}
For any $\gamma$, $c$ such that $\gamma> c$,
%
%
\begin{eqnarray}
\label{cs00}
\Prol( \tilde{\cS}_{\gamma} \leq\cS_{\gamma}< \infty)&=&1,
\\
\label{cs0}
\Prol(A_{\dmle_{\gamma}} \leq\gamma)&=&1,
\\
\label{cs}
\Expel\bigl[(M_{\dmle_{\gamma}})^{2}\bigr] &\leq&\gamma.
\end{eqnarray}
Moreover, if $c, \gamma\rightarrow\infty$ so that $c=o(\gamma)$, then
%
%
\begin{equation}
\label{sat} \limsup_{\gamma\rightarrow\infty} \bigl(\Expel[A_{\tilde{\cS
}_{\gamma
}}] -\gamma\bigr)
\leq0.
\end{equation}
\end{lemma}
\begin{pf}
From the first inequality in (\ref{dist2}) we have $\tA\geq A-c$, therefore,
%
%
\begin{equation}
\label{path} \tilde{\cS}_{\gamma} \leq\inf\{ t \geq0\dvtx A_{t} -c
\geq\gamma-c \} = \cS_{\gamma}.
\end{equation}
From this inequality and (\ref{no}) we obtain (\ref{cs00}). Moreover,
since $A$ is the quadratic variation of $B$,
it has continuous and increasing paths, thus, from (\ref{cs00}) we
obtain $\Prol( A_{\tilde{\cS}_{\gamma}} \leq A_{\cS_{\gamma}}=\gamma)=1$.
Finally, from (\ref{wald2}) and (\ref{cs0}) we obtain
\[
\Expel\bigl[(M_{\dmle_{\gamma}})^{2}\bigr] = \Expel[A_{\dmle_{\gamma}}]
\leq\gamma,
\]
which proves (\ref{cs}) and implies (\ref{sat}).
\end{pf}

In the following theorem we show that $\tmle_{\dmle_{\gamma}}$ is a
consistent estimator of $\lambda$, even under an asymptotically low
communication rate.
%
%
\begin{theorem} \label{theo1}
$\Prol(\tmle_{\dmle_{\gamma}} \rightarrow\lambda)=1$ and $\Expel
[(\tmle_{\dmle_{\gamma}}- \lambda)^{2}] \rightarrow0$ as $\gamma, c,
\Delta
\rightarrow\infty$ so that $c, \Delta=o(\gamma)$.
\end{theorem}
\begin{pf}
Recalling from (\ref{score0}) that $B=\lambda A +M$, we have $\Prol$-a.s.
\begin{eqnarray*}
\tmle_{\dmle_{\gamma}} &=& \biggl( \frac{\tB}{\tA} \biggr)_{\dmle
_{\gamma}} =
\biggl( \frac{\tB-B}{\tA} \biggr)_{\dmle_{\gamma}} + \biggl( \frac
{B}{\tA}
\biggr)_{\dmle_{\gamma}}
\\
&=& \biggl( \frac{ \tilde{B}-B}{\tilde{A}} \biggr)_{\dmle_{\gamma}} +
\lambda\biggl(
\frac{A}{\tA} \biggr)_{\dmle_{\gamma}} + \biggl( \frac
{M}{\tA}
\biggr)_{\dmle_{\gamma}}
\end{eqnarray*}
and, consequently,
%
%
\begin{equation}
\label{repre3} \tmle_{\dmle_{\gamma}} -\lambda= \biggl( \frac{ \tilde
{B}-B}{\tilde{A}}
\biggr)_{\dmle_{\gamma}} + \lambda\biggl( \frac{A-\tA}{\tA} \biggr
)_{\dmle_{\gamma}}
+ \biggl( \frac{M}{\tA} \biggr)_{\dmle_{\gamma}}.
\end{equation}
From the definition of $\dmle_{\gamma}$ it follows that $\tA_{\dmle
_{\gamma}} \geq\gamma-c$, whereas from (\ref{dist2}) we have
$|\tilde{B}-B|_{\dmle_{\gamma}}\leq\D$ and $(A-\tilde{A})_{\dmle
_{\gamma}} \leq c$. Therefore, 
%
%
\begin{equation}
\label{inemse} |\tmle_{\dmle_{\gamma}}- \lambda| \leq\frac{\D+ |\lambda|
c}{\gamma
-c} +
\frac{|M_{\dmle_{\gamma}}|}{\gamma-c}.
\end{equation}
The first term in the right-hand side clearly goes to 0 as $c, \D,
\gamma\rightarrow\infty$ so that $c, \D=o(\gamma)$. Moreover,
from (\ref{dds}) and (\ref{cs0}) we have $\Prol$-a.s.
%
%
\begin{equation}
\label{inemse2} \frac{|M_{\dmle_{\gamma}}|}{\gamma-c} = \frac
{|W_{A_{\dmle
_{\gamma
}}}|}{A_{\dmle_{\gamma}}} \frac{A_{\dmle_{\gamma}}}{\gamma-c} \leq
\frac{|W_{A_{\dmle_{\gamma}}}|}{A_{\dmle_{\gamma}}} \frac
{\gamma
}{\gamma-c}.
\end{equation}
If $c, \gamma\rightarrow\infty$ so that $c= o(\gamma)$, $\Prol
(A_{\dmle_{\gamma}} \rightarrow\infty)=1$, due to assumption (\ref{equA3}).
Therefore, the strong law of large numbers implies that the right-hand
side in (\ref{inemse2}) converges to 0 and, consequently,
$\Prol(\tmle_{\dmle_{\gamma}} \rightarrow\lambda)=1$ as $c,
\gamma
\rightarrow\infty$ so that $c= o(\gamma)$.

Moreover, if we square both sides in (\ref{inemse}), apply the
algebraic inequality $(x+y)^{2}\leq2(x^{2}+y^{2})$,
take expectations and use (\ref{cs}), we obtain
\[
\Expel\bigl[(\tmle_{\dmle_{\gamma}}- \lambda)^{2}\bigr] \leq2 \biggl(
\frac
{\Delta+ |\lambda| c}{\gamma-c} \biggr)^{2} + 2 \frac{ \gamma}{(\gamma-c)^{2}},
\]
which implies that $\Expel[(\tmle_{\dmle_{\gamma}}- \lambda)^{2}]
\rightarrow0$ as $ c, \D, \gamma\rightarrow\infty$ so that $c, \D
=o(\gamma)$.
\end{pf}

The consistency of $\tmle_{\dmle_{\gamma}}$ was established without any
additional conditions on the dynamics of the sensor processes. However,
it is clear that the suggested estimator cannot be asymptotically
efficient in such a general setup, since it does not have
any access to sufficient statistics of the form $A^{ij}$ with $(i,j)
\in\cD$.

Nevertheless, if every $A^{ij}$ with $i \neq j$ is deterministic, then
$\cD= \varnothing$ and the fusion center has access to all sufficient
statistics for $\lambda$. In this case, we can obtain an asymptotically
sharp lower bound for $A_{\dmle_{\gamma}}$, the observed Fisher
information that is utilized by the proposed estimator, which allows us
to establish its asymptotic optimality even under an asymptotically low
communication rate.
%
%
\begin{lemma}
If $\cD= \varnothing$, then $\tA_{t} \leq A_{t}$ for every $t \geq0$.
Consequently, for every $\gamma, c$ such that $\gamma>c$,
%
%
\begin{eqnarray}
\label{cs20}
\Prol(A_{\dmle_{\gamma}} \geq\gamma- c)&=&1,
\\
\label{cs2}
\Expel\bigl[ (M_{\mle_{\gamma}}- M_{\dmle_{\gamma}})^{2}\bigr] &\leq& c.
\end{eqnarray}
\end{lemma}
\begin{pf}
If $\cD= \varnothing$, then $d_{i}=0$ for every $1 \leq i \leq K$,
thus, from (\ref{iso}), (\ref{tA}) and the first inequality in (\ref
{dist}) we obtain
\[
\tA= \sum_{i=1}^{K} \tA^{i} +
\sum_{1 \leq j \neq i \leq K} A^{ij} \leq\sum
_{i=1}^{K} A^{i} + \sum
_{1 \leq j \neq i \leq K} A^{ij} = A.
\]
Then,\vspace*{1pt} from the definition of $\dmle_{\gamma}$ we have
$\Prol(A_{\dmle_{\gamma}} \geq\tA_{\dmle_{\gamma}} \geq\gamma-
c)=1$, which proves (\ref{cs20}). Finally, from (\ref{wald2}), (\ref
{path}) and (\ref{cs20}) we obtain
\[
\Expel\bigl[ (M_{\mle_{\gamma}}- M_{\dmle_{\gamma}})^{2}\bigr] = \Expel
[A_{\mle
_{\gamma}}- A_{\dmle_{\gamma}}] = \Expel[\gamma- A_{\dmle_{\gamma}}]
\leq c,
\]
which completes the proof.
\end{pf}
%
%
\begin{theorem} \label{theo2}
If $\cD= \varnothing$, then $ \sqrt{\gamma} ( \tmle_{\dmle
_{\gamma}}-
\lambda) \rightarrow\cN(0,1)$ as $ c, \Delta, \gamma\rightarrow
\infty$ so that
$c, \Delta=o(\sqrt{\gamma})$.
\end{theorem}
\begin{pf}
Since $\sqrt{\gamma} (\hat{\lambda}_{\mle_{\gamma}}-\lambda)
\sim\cN
(0,1)$ for every $\gamma>0$, it suffices to show that $\sqrt{\gamma}
|\tmle_{\dmle_{\gamma}} - \hat{\lambda}_{\mle_{\gamma}}|$
converges to
zero in probability
as $\gamma, c, \Delta\rightarrow\infty$ so that $c, \Delta
=o(\sqrt{\gamma})$. 
Indeed, from (\ref{deco2}) and (\ref{repre3}) we have $\Prol$-a.s.
\[
\tmle_{\dmle_{\gamma}} - \hat{\lambda}_{\mle_{\gamma}} = \biggl( \frac
{\tB-B} {\tA}
\biggr)_{\dmle_{\gamma}} + \lambda\biggl( \frac{A-\tA}{\tA} \biggr
)_{\dmle_{\gamma}}
+ \biggl( \frac{M}{\tA} \biggr)_{\dmle_{\gamma}} - \biggl( \frac{M}{A}
\biggr)_{\mle_{\gamma}}.
\]
Since $\tA_{\dmle_{\gamma}} \geq\gamma- c$ and from (\ref{dist2}) we
have $|\tilde{B}-B|_{\dmle_{\gamma}}\leq\D$ and $(A-\tilde
{A})_{\dmle
_{\gamma}} \leq c$,
%
%
\begin{equation}
\label{ww} \sqrt{\gamma} |\tmle_{\dmle_{\gamma}} - \hat{\lambda}_{\mle
_{\gamma}}|
\leq\sqrt{\gamma} \frac{\D+ |\lambda| c}{\gamma- c} + \sqrt{\gamma} \biggl|
\biggl(
\frac{M}{\tA} \biggr)_{\dmle_{\gamma}} - \biggl( \frac
{M}{A}
\biggr)_{\mle_{\gamma}} \biggr|.
\end{equation}
The first term in the right-hand side of (\ref{ww}) converges to 0 as
$c, \D, \gamma\rightarrow\infty$ so that $c,\D=o(\sqrt{\gamma})$.
Moreover, since $A_{\mle_{\gamma}}=\gamma$ and $\tA_{\dmle_{\gamma}}
\geq\gamma- c$,
%
%
\begin{eqnarray}\label{www}
\sqrt{\gamma} \biggl| \biggl( \frac{M}{\tA} \biggr)_{\dmle
_{\gamma}} - \biggl(
\frac{M}{A} \biggr)_{\mle_{\gamma}} \biggr| &=& \sqrt{\gamma} \biggl| \biggl(
\frac{M}{\tA} \biggr)_{\dmle
_{\gamma}} - \frac{M_{\dmle_{\gamma}}}{\gamma} +
\frac{M_{\dmle_{\gamma}}}{\gamma} - \frac{M_{\mle_{\gamma
}}}{\gamma} \biggr|
\nonumber
\\
&\leq& \frac{1}{\sqrt{\gamma}} \biggl[ | M_{\dmle_{\gamma}}| \frac
{\gamma- \tA_{\dmle_{\gamma}}}{\tA_{\dmle_{\gamma}}} +
|M_{\dmle_{\gamma}} - M_{\mle_{\gamma}}| \biggr]
\\
&\leq& \frac{1}{\sqrt{\gamma}} \biggl[ |M_{\dmle_{\gamma}}| \frac
{c}{\gamma-c} + |
M_{\dmle_{\gamma}} - M_{\mle_{\gamma}}| \biggr].\nonumber
\end{eqnarray}
From the Cauchy--Schwarz inequality, (\ref{cs}) and (\ref{cs2}) we have
\begin{eqnarray*}
\Expel\bigl[|M_{\dmle_{\gamma}}|\bigr] &\leq& \sqrt{\Expel\bigl[M^{2}_{\dmle
_{\gamma
}}
\bigr]} \leq\sqrt{\gamma},
\\
\Expel\bigl[|M_{\dmle_{\gamma}} - M_{\mle_{\gamma}}|\bigr] &\leq& \sqrt{\Expel
\bigl[(M_{\dmle_{\gamma}} - M_{\mle_{\gamma}})^{2}\bigr]} \leq\sqrt{c}.
\end{eqnarray*}
Then, taking expectations in (\ref{www}), we obtain
\[
\sqrt{\gamma} \Expel\biggl[ \biggl| \biggl( \frac{M}{\tA} \biggr)_{\dmle
_{\gamma}}
- \biggl( \frac{M}{A} \biggr)_{\mle_{\gamma}} \biggr| \biggr] \leq
\frac{c}{\gamma-c} + \sqrt{\frac{c}{\gamma}}.
\]
Therefore, the second term in the right-hand side of (\ref{ww})
converges to 0 in probability, due to Markov's inequality,
as $c, \D, \gamma\rightarrow\infty$ so that $c=o(\gamma)$. This
concludes the proof.
\end{pf}
%
%
\begin{corollary} \label{cor}
If $\cD= \varnothing$, then $(\dmle_{\gamma}, \tmle_{\dmle
_{\gamma}})$
is asymptotically optimal as $\gamma, c, \Delta\rightarrow\infty$
so that
$c, \Delta=o(\sqrt{\gamma})$.
\end{corollary}
\begin{pf}
This is a consequence of (\ref{sat}) and Theorem~\ref{theo2}.
\end{pf}

\subsection{Remarks and examples}\label{sec3.3}
For the implementation of the proposed estimator, the fusion center
does not need to record the values of the communication times. It
simply needs to
keep track of $\tB^{1},\ldots, \tB^{K}$ and---if necessary---$\tA
^{1},\ldots, \tA^{K}$, and update them whenever it receives a relevant
message. Since these statistics are defined recursively, at most $2K$
values need to be stored at any given time.

Theorems~\ref{theo0},~\ref{theo1} and~\ref{theo2} remain valid if $c$
and $\D$ are held fixed as $t \rightarrow\infty$ or $\gamma
\rightarrow\infty$.
Moreover, they remain valid if we use in the definitions of $\tau
_{n}^{i,B}$ and $\tau_{n}^{i,A}$ time-varying, positive thresholds,
$\overline{\Delta}{}^{i}_{n}$, $\underline{\Delta}^{i}_{n}$,
$c_{n}^{i}$, so that
\[
\overline{\Delta}{}^{i}_{n} \leq\overline{
\Delta}{}^{i},\qquad \underline{\Delta}^{i}_{n} \leq
\underline{\Delta}^{i},\qquad c_{n}^{i} \leq
c^{i} \qquad\forall n \in\mathbb{N}.
\]
Therefore, it may be possible to improve the performance of the
proposed estimator by introducing linear or curved boundaries and
optimizing over the additional parameters.

We close this section with some examples that illustrate our main
results. Thus,
let $\sigma_{t}:= [\sigma^{ij}_{t}]$ be an $\{\cFt\}$-adapted, square
matrix of size K,
set $\alpha_{t}:=\sigma_{t} \sigma'_{t}$, where $\sigma'_{t}$ is the
transpose of $\sigma_{t}$,
and consider the following\vadjust{\goodbreak} special case of model (\ref{model}):
%
%
\begin{equation}
\label{model5}\quad Y_{t}^{i}= \lambda\sum
_{j=1}^{K} \int_{0}^{t}
X_{s}^{j} \alpha^{ij}_{s} \,\ddd s + \sum
_{j=1}^{K} \int_{0}^{t}
\sigma^{ij}_{s} \,\ddd W^{j}_{s},\qquad t \geq0,
1 \leq i \leq K,
\end{equation}
where $(W^{1},\ldots, W^{K})$ is a $K$-dimensional $\Prol$-Brownian
motion. The observed Fisher information $\{A_{t}\}$ then becomes
%
%
\begin{equation}
\label{detfisher2} A_{t} = \sum_{i=1}^{K}
\sum_{j=1}^{K} \int_{0}^{t}
X^{i}_{s} X^{j}_{s}
\alpha^{ij}_{s} \,\ddd s,\qquad t \geq0.
\end{equation}

In Theorem~\ref{theo0}, we stated the asymptotic properties of the
proposed estimator when $A_{t}$ is deterministic. This assumption is
clearly satisfied when
there are real functions $b_{i}, \rho_{ij}\dvtx[0, \infty) \rightarrow
\mathbb{R}$ so that $X_{t}^{i}= b_{i}(t)$ and $\alpha^{ij}_{t}= \rho
_{ij}(t)$ for every $1 \leq i,j \leq K$, in which case
%
%
\begin{equation}
\label{detfisher} A_{t} = \sum_{i=1}^{K}
\sum_{j=1}^{K} \int_{0}^{t}
b_{i}(s) b_{j}(s) \rho_{ij}(s) \,\ddd s,\qquad t \geq0,
\end{equation}
and $(Y^{1},\ldots, Y^{K})$ is a Gaussian process with independent
increments. However, Theorem~\ref{theo0} also applies when $X_{t}^{i}=
b_{i}(t) / Y_{t}^{i}$ and $\alpha^{ij}_{t}= \rho_{ij}(t) Y_{t}^{i}
Y_{t}^{j}$,
in which case A is still given by (\ref{detfisher}).

In Theorem~\ref{theo2}, we proved that the proposed estimator is
asymptotically optimal when $A^{ij}$ is deterministic for every $i \neq
j$. This condition is clearly satisfied when $\sigma^{ij}=0$ for every
$ i\neq j$, in which case $Y^{1},\ldots, Y^{K}$ are independent,
$\alpha^{ii}= (\sigma^{ii})^{2}$ and (\ref{model5}), (\ref
{detfisher2}) become
\begin{eqnarray*}
Y_{t}^{i} &=& \lambda\int_{0}^{t}
X_{s}^{i} \alpha^{ii}_{s} \,\ddd s + \int
_{0}^{t} \sqrt{\alpha^{ii}_{s}}
\,\ddd W^{i}_{s},\qquad t \geq0,
\\
A_{t} &=& \sum_{i=1}^{K} \int
_{0}^{t} \bigl(X_{s}^{i}
\bigr)^{2} \alpha^{ii}_{s} \,\ddd s,\qquad t \geq0.
\end{eqnarray*}
If, in particular, $X^{i}$ is a nonzero constant and $\alpha
^{ii}=Y^{i}$, then $Y^{i}$ is a square-root diffusion,
whereas if $X^{i}=Y^{i}$ and $\alpha^{ii}$ is a positive constant, then
$Y^{i}$ is an Ornstein--Uhlenbeck process.

\section{The Brownian case}\label{sec4}
In this section we assume that $\langle Y^{i},Y^{j}\rangle_{t}=0$,
$\langle Y^{i},Y^{i} \rangle_{t}=t$ and $X^{i}_{t}=x_{i}$,
where $x_{i} \neq0$ is a known constant, for every $1 \leq i \neq j
\leq K$ and $t \geq0$.

Thus, $B_{t}^{i}= x_{i} Y_{t}^{i}$, $A_{t}^{i}= (x_{i})^{2} t$,
$A_{t}= \sum_{i=1}^{K} A_{t}^{i}$ and (\ref{model}) reduces to
\[
Y_{t}^{i}= \lambda x_{i} t + N_{t}^{i},\qquad
t \geq0, i=1,\ldots, K,
\]
where $N^{1},\ldots, N^{K}$ are independent, standard Brownian motions
under $\Prol$.

Since the filtrations $\{\cFt^{1}\},\ldots, \{\cFt^{K}\}$ are
independent, for every $1 \leq i \leq K$ and $t > 0$ we have
%
%
\begin{equation}
\label{probi} \frac{\mathrm{d} \Prol}{\mathrm{d} \Pro_{0}} \bigg|_{\cFt^{i}}
= e^{\lambda B_{t}^{i}- (\lambda^2 /2) A_{t}^{i}}=
e^{\lambda
B_{t}^{i}- (\lambda x_{i})^2 t/2}.
\end{equation}
We also assume, for simplicity, that $\overline{\Delta
}{}^{i}=\underline{\Delta}^{i}=\Delta^{i}$ for every
$1\leq i
\leq K$, thus, $\D=\sum_{i=1}^{K} \Delta^{i}$ and
%
%
\begin{eqnarray}
\label{t} \tau_{n}^{i,B} &=& \inf\bigl\{t \geq
\tau_{n-1}^{i,B}\dvtx \bigl|B^{i}_{t} -
B^{i}_{\tau^{i,B}_{n-1}}\bigr| \geq\Delta^{i}\bigr\},
\\
\label{z} z_n^i &=& \cases{1, &\quad if $B^{i}_{\tau_{n}^{i,B}}-
B^{i}_{\tau
_{n-1}^{i,B}} \geq\Delta^{i}$,
\vspace*{2pt}\cr
0, &\quad if
$B^{i}_{\tau_{n}^{i,B}}- B^{i}_{\tau_{n-1}^{i,B}} \leq-
\Delta^{i}$.}
\end{eqnarray}
We denote by $\delta_{n}^{i}$ the time between the arrival of the
$(n-1)$th and the $n$th message from sensor $i$ and by $m_{t}^{i}$ the
number of transmitted messages by sensor $i$ up to time~$t$, that is,
%
%
\begin{equation}
\label{renew} \delta_{n}^{i}:=\tau_{n}^{i,B}-
\tau_{n-1}^{i,B},\qquad m_{t}^{i}:=\max\bigl\{n
\in\mathbb{N}\dvtx \tau_{n}^{i} \leq t \bigr\}.
\end{equation}
Since $\{A_{t}\}$ is deterministic, $\tau_{n}^{i,A}=\infty$ for every
$1 \leq i \leq K$ and $n \in\mathbb{N}$ and the fusion center
filtration becomes
\[
\tilde{\cFt}= \sigma\bigl(\delta_{n}^{i},
z_{n}^{i}; n \leq m_{t}^{i}, 1\leq i
\leq K\bigr),\qquad t \geq0.
\]
Moreover, $\tA:=A$ and $\tA^{i}:=A^{i}$ for every $i$, however, we now
define the following $\{\tilde{\cFt}\}$-adapted statistics:
%
%
\begin{equation}
\label{cAi} \check{A}_{t}^{i}:= |x_{i}|^{2} \sum_{j=1}^{m_{t}^{i}}
\delta_{j}^{i},\qquad \check{A}_{t}:= \sum_{i=1}^{K} \check{A}_{t}^{i},\qquad t
\geq0.
\end{equation}
That is, $\check{A}_{t}$ is an approximation of $A_{t}$ that relies
only on the communication times $\{\tau_{n}^{i,B}; n \leq m_{t}^{i},
1 \leq i \leq K\}$.

Since Brownian motion ``restarts'' at stopping times, each $(\delta
_{n}^{i},z_{n}^{i})_{n \in\mathbb{N}}$ is a sequence of i.i.d. pairs, thus,
each $(m_{t}^{i})_{t \geq0}$ is a renewal process. Moreover, it is
possible to obtain a series representation for the joint density of the
pair $(\delta_{1}^{i},z_{1}^{i})$ under~$\Prol$,
\[
\bar{p}_{i}(t; \lambda):= \frac{\Prol(\delta_{1}^{i} \in \ddd t,
z_{1}^{i}=1)}{\ddd t},\qquad \uunderline{p}_{i}(t; \lambda):=
\frac{\Prol(\delta_{1}^{i} \in \ddd t, z_{1}^{i}=0)}{\ddd t}.
\]
This representation is the content of the following lemma, for which we
need to define the following functions:
\[
g(t; x):= \sum_{n=-\infty}^{\infty} h\bigl(t;
(4n+1)x\bigr),\qquad h(t;x):= \frac{x}{\sqrt{2 \pi t^{3}}} e^{-x^{2}/2t},\qquad
t,x \geq0.
\]

%
\begin{lemma} \label{p1}
For every $1 \leq i \leq K$ and $t>0$,
\begin{eqnarray*}
\bar{p}_{i}(t; \lambda)&=&e^{\lambda\Delta^{i}- 0.5 (\lambda
x_{i})^{2} t
} g \bigl(t;
\Delta^{i}/ |x_{i}|\bigr),
\\
\uunderline{p}_{i}(t; \lambda) &=& e^{-\lambda\Delta^{i}-0.5
(\lambda
x_{i})^{2} t} g \bigl(t;
\Delta^{i}/ |x_{i}|\bigr).
\end{eqnarray*}
\end{lemma}
\begin{pf}
From (\ref{t}) and (\ref{renew}) we have
%
%
\begin{equation}
\label{delt} \delta_{1}^{i} = \inf\bigl\{t \geq0\dvtx
\bigl|Y^{i}_{t}\bigr| \geq\Delta^{i}/ |x_{i}|
\bigr\},\qquad n \in\mathbb{N}.
\end{equation}
Since $Y^{i}$ is a standard Brownian motion under $\Pro_{0}$, it is
well known (see, e.g.,~\cite{kar}, page 99) that $\bar{p}_{i}(t; 0)=
\uunderline{p}_{i}(t;0)= g(t; \Delta^{i}/|x_{i}|)$. Then,\vspace*{1pt}
changing the measure $\Prol\mapsto\Pro_{0}$ (similarly, e.g., to
\cite{kar}, page 196), we obtain the desired result.
\end{pf}

The following lemma describes some properties of the communication scheme
that remain valid in the case of discrete sampling at the sensors,
which we treat in Section~\ref{sec4.2}. In order to lighten the
notation, we
denote by $\Theta(\Delta^{i})$ a term that when divided by $\Delta^{i}$ is
asymptotically bounded from above and below as $\Delta^{i}\rightarrow
\infty$.
%
%
\begin{lemma} \label{p2}
\textup{(a)} For any $t, \Delta^{i}>0$,
%
%
\begin{eqnarray}
\label{lor1}
\Expel\Biggl[\sum_{j=1}^{m_{t}^{i}+1}
\delta_{j}^{i}- t \Biggr] &\leq& \frac{\Expel[(\delta
_{1}^{i})^{2}]}{\Expel[\delta_{1}^{i}]},
\\
\label{lor2}
\Expel\Biggl[t- \sum_{j=1}^{m_{t}^{i}}
\delta_{j}^{i} \Biggr] &\leq& \frac{\Expel[(\delta_{1}^{i})^{2}]}{\Expel
[\delta_{1}^{i}]}.
\end{eqnarray}

\textup{(b)} As $t, \Delta^{i}\rightarrow\infty$,
%
%
\begin{eqnarray}
\label{rates}
&\displaystyle \Expel\bigl[\delta_{1}^{i}\bigr] = \Theta\bigl(
\Delta^{i}\bigr),\qquad \Var_{\lambda
}\bigl[\delta_{1}^{i}
\bigr]= \Theta\bigl(\Delta^{i}\bigr),&
\\
\label{calos2}
&\displaystyle 0 \leq\Expel\bigl[A^{i}_{t}- \check{A}^{i}_{t}
\bigr] \leq\Theta\bigl(\Delta^{i}\bigr),&
\\
\label{calos}
&\displaystyle \Expel\bigl[m_{t}^{i}\bigr] \leq t/ \Theta\bigl(
\Delta^{i}\bigr) + 1/ \Theta\bigl(\Delta^{i}\bigr).&
\end{eqnarray}
\end{lemma}
\begin{pf}
(a) Since $(\delta_{n}^{i})_{n \in\mathbb{N}}$ is a sequence of i.i.d.
random variables, (\ref{lor1}) follows from Theorem 1 in Lorden
\cite{lorden} and (\ref{lor2}) from Lorden~\cite{lorden}, page 526.

(b) Recall from (\ref{delt}) that $\delta_{1}^{i}$ is the first time a
Brownian motion with drift $\lambda x_{i}$ exits the symmetric interval
$(-\Delta^{i}/|x_{i}|, \Delta^{i}/|x_{i}|)$. Then, as $\Delta
^{i}\rightarrow\infty$, from
Wald's identity we have
%
%
\begin{equation}\label{moments0}
\Expel\bigl[\delta_{1}^{i}\bigr] = \frac{\Delta^{i}/ |x_{i}|}{|\lambda
x_{i}|}
\bigl(1+o(1)\bigr),
\end{equation}
whereas from Martinsek~\cite{mar} we have
%
%
\begin{equation}\label{moments}
\Var_{\lambda}\bigl[\delta_{1}^{i}\bigr] =
\frac{\Delta^{i}/|x_{i}|}{|\lambda
x_{i}|^{3}} \bigl(1+o(1)\bigr).
\end{equation}
Then, from (\ref{moments0}) and (\ref{moments}) we obtain (\ref{rates}),
whereas from (\ref{cAi}), (\ref{lor2}) and (\ref{rates}) we obtain
(\ref
{calos2}).

Finally, since $m_{t}^{i}+1$ is a stopping time with respect to the
filtration generated by the pairs $(\delta_{n}^{i},z_{n}^{i})_{n \in
\mathbb{N}}$, from Wald's identity and (\ref{lor1}) we have
\[
\Expel\bigl[m_{t}^{i}+1\bigr] \Expel\bigl[
\delta_{1}^{i}\bigr] = \Expel\Biggl[ \sum
_{j=1}^{m_{t}^{i}+1} \delta_{j}^{i} \Biggr]
\leq t+ \frac{\Expel[(\delta^{i}_{1})^{2}]}{\Expel[\delta_{1}^{i}]}
\]
and, consequently,
\[
\Expel\bigl[m_{t}^{i}\bigr] \leq\frac{t}{\Expel[\delta_{1}^{i}]} +
\frac
{\Var_{\lambda}[\delta^{i}_{1}]}{(\Expel[\delta_{1}^{i}])^{2}}.
\]
From this inequality and (\ref{rates}) we obtain (\ref{calos}), which
completes the proof.~%
\end{pf}

\subsection{Likelihood-based estimation at the fusion center}\label{sec4.1}
Let $\tilde{\cL}_{t} (\lambda)$ and $\tilde{\ell}_{t}(\lambda)$
be the
likelihood and the log-likelihood function of $\lambda$ that correspond
to $\tilde{\cF}_{t}$, the accumulated information at the fusion center
up to time $t$. The following proposition describes the structure of
the corresponding score function.
%
%
\begin{proposition} For any $ t > 0$,
%
%
\begin{equation}
\label{score} \frac{\mathrm{d} \tilde{\ell}_{t}(\lambda)}{\mathrm{d}
\lambda
} = \Biggl\{ \sum_{i=1}^{K}
\Expel\bigl[ B^{i}_{t} | m_{t}^{i} \bigr]
-\lambda A_{t} \Biggr\} + \{ \tB_{t} - \lambda
\check{A}_{t} \}.
\end{equation}
\end{proposition}
\begin{pf}
Suppose that $m_{t}^{i}=m_{i}$, that is, sensor $i$ has transmitted
$m_{i}$ messages to the fusion center up to time $t$,
where $m_{i}$ is some nonnegative integer. Then, since all pairs $\{
(z_{n}^{i}, \delta_{n}^{i}), n \in\mathbb{N}, 1 \leq i \leq K\}$
are independent,
the fusion likelihood function has the following form:
\[
\tilde{\cL}_{t} (\lambda):= \prod_{i=1}^{K}
\Pro_{\lambda
}\bigl(m_{t}^{i}=m_{i}\bigr)
\Biggl( \prod_{n=1}^{m_{i}}
\bar{p}_{i}\bigl( \delta_{n}^{i};\lambda
\bigr)^{z_{n}^{i}} \cdot\uunderline{p}_{i}\bigl(\delta_{n}^{i};
\lambda\bigr)^{1-z_{n}^{i}} \Biggr)^{\ind{m_{i}>0}}.
\]
Due to Lemma~\ref{p1}, the corresponding log-likelihood function becomes
\begin{eqnarray*}
\tilde{\ell}_{t}(\lambda) &=& \sum_{i=1}^{K}
\log\Pro_{\lambda
}\bigl(m_{t}^{i}=m_{i}\bigr)
\\
&&{} + \sum_{i=1}^{K} \ind{m_{i}>0}
\sum_{n=1}^{m_{i}} \biggl[ \lambda
\Delta^{i}-\frac{(\lambda x_{i} )^{2} \delta_{n}^{i}}{2} + \log g\bigl
(\delta_{n}^{i};
\Delta^{i}/|x_{i}|\bigr) \biggr] z_{n}^{i}
\nonumber
\\
&&{} + \sum_{i=1}^{K} \ind{m_{i}>0}
\sum_{n=1}^{m_{i}} \biggl[ -\lambda
\Delta^{i}- \frac{(\lambda x_{i} )^{2} \delta_{n}^{i}}{2} + \log g\bigl
(\delta_{n}^{i};
\Delta^{i}/|x_{i}|\bigr) \biggr] \bigl(1-z_{n}^{i}
\bigr).
\end{eqnarray*}
Then, recalling the definition of $\tB$ in (\ref{tBi})--(\ref{tB}) and
of $\check{A}$ in (\ref{cAi}),
\[
\frac{\mathrm{d} \tilde{\ell}_{t}(\lambda) }{\mathrm{d} \lambda} = \sum
_{i=1}^{K}
\frac{\mathrm{d}}{\mathrm{d} \lambda} \bigl( \log\Pro_{\lambda}\bigl
(m_{t}^{i}=m_{i}
\bigr) \bigr) +\tB_{t} - \lambda\check{A}_{t}.
\]
Since $\{m_{t}^{i}=m_{i} \} \in\cFt^{i}$, changing the measure $\Prol
\mapsto\Pro_{0}$, we have
\[
\Pro_{\lambda}\bigl( m_{t}^{i}=m_{i}\bigr) =
\Exp_{0} \bigl[e^{\lambda
B^{i}_{t} - {\lambda^{2}}A^{i}_{t}/2} \ind{ m_{t}^{i}=m_{i}}
\bigr]
\]
and, consequently,
\begin{eqnarray*}
\frac{\mathrm{d}}{\mathrm{d} \lambda} \bigl( \log\Pro_{\lambda}\bigl(
m_{t}^{i}=m_{i}
\bigr) \bigr) &=& \frac{\Exp_{0} [e^{\lambda B^{i}_{t} - {\lambda^{2}}
A^{i}_{t}/{2}} (B^{i}_{t} -\lambda A^{i}_{t})
\ind{m_{t}^{i}=m_{i}} ]}{\Pro_{\lambda}( m_{t}^{i}=m_{i})}
\\
&=& \frac{\Expel[B^{i}_{t} \ind{m_{t}^{i}=m_{i}}]- \lambda
A^{i}_{t} \Pro_{\lambda}( m_{t}^{i}=m_{i})}{\Pro_{\lambda}(
m_{t}^{i}=m_{i})}
\\
&=& \Expel\bigl[ B^{i}_{t} | m_{t}^{i}=m_{i}
\bigr] - \lambda A^{i}_{t},
\end{eqnarray*}
which implies (\ref{score}).
\end{pf}

Note that the second term in (\ref{score}) reflects the information
from the communication times and the transmitted messages, whereas the
first term reflects the information between transmissions.

At time $t$, the fusion center should ideally estimate $\lambda$ with
the fusion center MLE, that is, the root of the score function (\ref{score}).
However, since $\Expel[ B^{i}_{t} | m_{t}^{i}]$ does not admit a
simple, closed-form expression as a function of $\lambda$, we can only
approximate
this conditional expectation and obtain an \textit{approximate} fusion
center MLE.

If we replace each $\Expel[ B^{i}_{t} | m_{t}^{i} ]$ with the
corresponding unconditional expectation,
$\Expel[ B^{i}_{t}] = \lambda A^{i}_{t}$, the first term in (\ref
{score}) vanishes and we obtain the following estimator:
%
%
\begin{equation}
\label{checkmle} \check{\lambda}_{t}:= \frac{\tB_{t}}{\check{A}_{t}},\qquad
t \geq \min_{1 \leq i \leq K} \tau_{1}^{i}.
\end{equation}
On the other hand, if we approximate $\Expel[ B^{i}_{t} | m_{t}^{i} ]$
with $\lambda\check{A}^{i}_{t}$,
we recover the estimator $\{\tilde{\lambda}_{t}\}$ that was defined in
(\ref{dbmled}) and whose asymptotic properties were established in
Theorem~\ref{theo0}.
In the following proposition we show that, in the special Brownian case
that we consider in this section,
$\check{\lambda}_{t}$ has similar asymptotic behavior as $\tilde
{\lambda}_{t}$.
%
%
\begin{proposition} \label{prop2}
If $t, \D\rightarrow\infty$ so that $\D=o(t)$, then $\check
{\lambda
}_{t}$ converges to $\lambda$ in probability. If additionally
$\D=o(\sqrt{t})$, then $\sqrt{A_{t}} (\check{\lambda}_{t} -
\lambda)
\rightarrow\cN(0,1)$, that is, $\check{\lambda}_{t}$ is an
asymptotically optimal estimator of $\lambda$.\vadjust{\goodbreak}
\end{proposition}
\begin{pf}
From the definition of $\tilde{\lambda}_t$ in (\ref{dbmled}) and
$\check
{\lambda}_{t}$ in (\ref{checkmle}) we have
%
%
\begin{equation}
\label{rep} \check{\lambda}_{t}- \tilde{\lambda}_{t} =
\frac{\tB_{t}}{\check
{A}_{t}}- \frac{\tB_{t}}{A_{t}} = \frac{A_{t}}{\check{A}_{t}} \frac
{A_{t}-\check{A}_{t}}{A_{t}}
\tilde{\lambda}_{t},\qquad t \geq0.
\end{equation}
From (\ref{calos2}) it follows that
%
%
\begin{equation}
\label{rep2} 0 \leq\frac{\Expel[A_{t}-\check{A}_{t}]}{A_{t}} = \frac
{1}{A_{t}} \sum
_{i=1}^{K} \Expel\bigl[A^{i}_{t}-
\check{A}^{i}_{t}\bigr] \leq\sum_{i=1}^{K}
\frac
{\Theta(\Delta^{i})}{A_{t}}= \frac{\Theta(\D)}{A_{t}}.
\end{equation}
Therefore, Markov's inequality implies that $(A_{t}-\check
{A}_{t})/A_{t}$ converges to 0 and $A_{t}/\check{A}_{t}$ converges to 1
in probability as $t, \D\rightarrow\infty$ so that $\D=o(t)$, since
$A_{t}$ is a linear function of $t$. Moreover, from Theorem~\ref{theo0}
we know that $\tilde{\lambda}_{t}$ converges to $\lambda$ in
probability if $\D=o(t)$. Thus, we conclude that $\check{\lambda}_{t}$
also converges to $\lambda$ in probability as $t, \D\rightarrow
\infty
$ so that $\D=o(t)$.

In order to prove that $\check{\lambda}_{t}$ is asymptotically optimal,
it suffices to show that $\sqrt{A_{t}} | \check{\lambda}_{t}- \tilde
{\lambda}_{t}|$ converges to 0 in probability as $t, \D\rightarrow
\infty$ so that $\D=o(\sqrt{t})$, which also follows from (\ref{rep})
and (\ref{rep2}).
\end{pf}

\subsection{The case of discrete sampling}\label{sec4.2}
We now assume that each sensor observes its underlying process only at
a sequence of discrete and equidistant times $\{nh, n \in\mathbb{N}\}
$, where $h>0$ is a common sampling period. Thus, in what follows,
$t=h, 2h, \ldots\,$. The goal is to examine the effect of discrete
sampling on the proposed estimating scheme.

First of all, we observe that the centralized estimator,
%
%
\begin{equation}
\label{discmle} \hat{\lambda}_{t}= \frac{B_{t}}{A_{t}}=
\frac{\sum_{i=1}^{K} x_{i}
Y_{t}^{i}}{\sum_{i=1}^{K} (x_{i})^{2} t }
\end{equation}
is not affected by the discrete sampling of the underlying processes
and (\ref{asydet}) remains valid, that is, $\sqrt{A_{t}} (\hat
{\lambda
}_{t} - \lambda) \sim\cN(0,1)$ for every $t=h, 2h, \ldots\,$.

Moreover, the pairs $(\delta_{n}^{i},z_{n}^{i})_{n \in\mathbb{N}}$
remain i.i.d. and Lemma~\ref{p2} still holds. On the other hand, Lemma
\ref
{p1} is no longer valid and there is not an explicit formula for the
density of the pair $(\delta_{1}^{i},z_{1}^{i})$. However, the main
difference in the case of discrete sampling is that at any time $\tau
_{n}^{i,B}$ the fusion center
learns whether $B^{i}$ increased or decreased by at least $\Delta^{i}$ since
$\tau_{n-1}^{i,B}$, but does not learn by how much exactly.
In other words, the fusion center does not learn the size of the
realized overshoots,
%
%
\begin{equation}\qquad
\eta_{n}^{i}:= \bigl(B^{i}_{\tau_{n}^{i,B}}-
B^{i}_{\tau_{n-1}^{i,B}} - \Delta^{i} \bigr)^{+} +
\bigl(B^{i}_{\tau_{n}^{i,B}}- B^{i}_{\tau_{n-1}^{i,B}} +
\Delta^{i}\bigr)^{-},\qquad n \in\mathbb{N}.
\end{equation}
As a result, the statistic $\tB^{i}$, defined in (\ref{tBi}), is no
longer equal to $B^{i}$ at the communication times $(\tau_{n}^{i,B})_{n
\in\mathbb{N}}$ and the distance $|B_{t}^{i}-\tB_{t}^{i}|$ is no
longer bounded by $\Delta^{i}$. Therefore, Theorem~\ref{theo0}, which
establishes the consistency and asymptotic optimality of the proposed
estimator, $\tilde{\lambda}_t= \tB_{t} / A_{t}$, under the assumption
of continuous-time sensor observations may not hold when the sensors
observe their underlying processes at discrete times.

Our goal is to determine under what conditions the consistency and
asymptotic optimality of $\tilde{\lambda}_{t}$ are preserved in the
context of discrete sampling at the sensors. In order to do so, we need
to estimate the inflicted performance loss due to the unobserved
overshoots. The following lemma is very useful in this direction.
%
%
\begin{lemma} \label{l}
For every $1 \leq i \leq K$,
%
%
\begin{equation}
\label{estim1} \bigl| B_{t}^{i} - \tB_{t}^{i}\bigr|
\leq\Delta^{i}+ \sum_{j=1}^{m_{t}^{i}}
\eta_{j}^{i},\qquad t \geq0,
\end{equation}
and the overshoots $(\eta_{n}^{i})_{n \in\mathbb{N}}$ are i.i.d. with
%
%
\begin{equation}
\label{estim} \sup_{\Delta^{i}>0} \Expel\bigl[\eta_{1}^{i}
\bigr] =\calo\bigl(\sqrt[3]{h}\bigr).
\end{equation}
\end{lemma}
\begin{pf}
For every $t \geq0$ we have
\begin{eqnarray*}
B_{t}^{i} - \tB_{t}^{i} &=&
B^{i}_{t} - B^{i}_{\tau^{i,B}_{m_{t}^{i}}} + \sum
_{j=1}^{m_{t}^{i}} \bigl(B^{i}_{\tau^{i,B}_{j}} -
B^{i}_{\tau
^{i,B}_{j-1}}\bigr) - \tB_{t}^{i}
\\
&=& B^{i}_{t} - B^{i}_{\tau^{i,B}_{m_{t}^{i}}} + \sum
_{j=1}^{m_{t}^{i}} \bigl[ \bigl(B^{i}_{\tau^{i,B}_{j}}
- B^{i}_{\tau^{i,B}_{j-1}}\bigr) - \bigl[\Delta^{i}
z_{j}^{i} - \Delta^{i}\bigl(1-z_{j}^{i}
\bigr) \bigr] \bigr],
\end{eqnarray*}
which implies (\ref{estim1}). It is obvious that the overshoots $(\eta
_{n}^{i})_{n \in\mathbb{{N}}}$ are i.i.d.
In order to prove (\ref{estim}), we write $\delta_{1}^{i}=\min\{
\underline{\delta}_{1}^{i}, \overline{\delta}{}^{i}_{1}\}$, where
\[
\underline{\delta}_1^i:=\inf\bigl\{nh\dvtx
B_{nh}^i \le-\Delta^{i}\bigr\},\qquad \overline{
\delta}{}^i_1:=\inf\bigl\{nh\dvtx B_{nh}^i
\ge\Delta^{i}\bigr\}.
\]
Then, from Theorem 3 of Lorden~\cite{lorden} it follows that for
\textit{any} $r \geq1$,
%
%
\begin{eqnarray}\label{lord}
\sup_{\Delta^{i}>0} \Expel\bigl[\eta_1^i\bigr] &\leq&
\max\bigl\{ \Expel\bigl[ B_{\overline
{\delta}{}^i_1}^i - \Delta^{i}
\bigr], -\Expel\bigl[B^{i}_{\underline{\delta}_1^i}+ \Delta^{i}\bigr]
\bigr\}
\nonumber\\[-8pt]\\[-8pt]
&\leq& \sqrt[r]{\frac{r+2}{r+1} \frac{\Expel
[|B^{i}_{h}|^{r+1}]}{|\Expel[B^{i}_{h}]|}}.\nonumber
\end{eqnarray}
Since $Y^{i}_{h} \sim\cN(\lambda x_{i} h, h)$ under $\Prol$ and
$B^{i}_{h}=x_{i} Y_{h}^{i}$,
\begin{eqnarray*}
\Expel\bigl[B^{i}_{h}\bigr] &=& \lambda
(x_{i})^{2} h,
\\
\Expel\bigl[\bigl(B^{i}_{h}\bigr)^{4}\bigr] &=&
(x_{i})^{4} \bigl[ (\lambda x_{i}
h)^{4} + 6 (\lambda x_{i} h)^{2} h+ 3
h^{2} \bigr]
\\
&=& 3 (x_{i})^{4} h^{2} \bigl(1+o(1)\bigr)
\qquad\mbox{as } h \rightarrow0.
\end{eqnarray*}
Setting $r=3$ in (\ref{lord}) completes the proof.\vadjust{\goodbreak}
\end{pf}

In the following theorem we show that $\tilde{\lambda}_{t}$ remains
consistent as $t \rightarrow\infty$ for any given, fixed sampling
period, $h>0$,
as long as the communication rate of every sensor is asymptotically low.
%
%
\begin{theorem} \label{theo3}
If $t, \Delta^{i}\rightarrow\infty$ so that $\Delta^{i}=o(t)$ for
every $1\leq i
\leq K$, then $\Expel[|\tilde{\lambda}_{t}-\lambda|] \rightarrow0$.
\end{theorem}
\begin{pf} Since $\Expel[|\hat{\lambda}_{t}-\lambda|] \rightarrow0$,
it suffices to show that $\Expel[|\tilde{\lambda}_{t}- \hat{\lambda
}_{t}|] \rightarrow0$. Indeed, from the definition of the two
estimators and (\ref{estim1}) we have
%
%
\begin{equation}
\label{usineq} |\tilde{\lambda}_{t}- \hat{\lambda}_{t}| \leq
\frac{1}{A_{t}} \sum_{i=1}^{K} \bigl|
\tB_{t}^{i}- B_{t}^{i}\bigr| \leq
\frac{\D}{A_{t}} + \frac{1}{A_{t}} \sum_{i=1}^{K}
\sum_{j=1}^{m_{t}^{i}+1} \eta_{j}^{i}.
\end{equation}
Since $m_{t}^{i}+1$ is a stopping time with respect to the filtration
generated by $(\delta_{n}^{i}, z_{n}^{i}, \eta_{n}^{i})_{n \in
\mathbb
{N}}$, from Wald's identity we obtain
%
%
\begin{equation}
\label{estim2} \Expel\Biggl[ \sum_{j=1}^{m_{t}^{i}+1}
\eta_{j}^{i} \Biggr]= \Expel\bigl[ \eta_{1}^{i}
\bigr] \Expel\bigl[ m_{t}^{i}+1\bigr].
\end{equation}
Taking expectations in (\ref{usineq}) and applying (\ref{estim2}), we obtain
%
%
\begin{equation}
\label{essence0} \Expel\bigl[|\tilde{\lambda}_{t}- \hat{
\lambda}_{t}|\bigr] \leq\frac{\D}{A_{t}} + \sum
_{i=1}^{K} \frac{ \Expel[ \eta_{1}^{i} ] \Expel[
m_{t}^{i}+1]}{A_{t}}.
\end{equation}
Then, from (\ref{calos}), (\ref{estim}) and the fact that $A_{t}$ is a
linear function of $t$ we have
%
%
\begin{equation}
\label{essence00} \Expel\bigl[|\tilde{\lambda}_{t}- \hat{
\lambda}_{t}|\bigr] \leq\frac
{\Theta(\D
)}{t} + \sum
_{i=1}^{K} \frac{\Expel[ \eta_{1}^{i} ]}{\Theta(\Delta^{i})}.
\end{equation}
If some $\Delta^{i}$ is fixed as $t \rightarrow\infty$, the second
term in
the right-hand side of (\ref{essence00}) does not go to 0 (unless $h
\rightarrow0$, in which case $\Expel[ \eta_{1}^{i}] \rightarrow0$ for
every $1 \leq i \leq K$, due to (\ref{estim})). However, if $\Delta^{i}
\rightarrow\infty$ so that $\Delta^{i}=o(t)$ for every $1 \leq i
\leq K$,
then both terms in the right-hand side of (\ref{essence00}) go to 0
for any given sampling period, $h>0$, which completes the proof.
\end{pf}

The proof of Theorem~\ref{theo3} suggests that the proposed estimator
is not consistent when both $\{\D^{i}, 1\leq i \leq K\}$
and $h$ are held fixed. In other words,
it is necessary to have either a high sampling rate $(h \rightarrow
0)$ in order to reduce the size of the unobserved overshoots
or a low communication rate in all sensors $(\Delta^{i}\rightarrow
\infty$ $
\forall1 \leq i \leq K)$ in order to reduce their accumulation rate.

However, an asymptotically low communication rate is not sufficient in
order to preserve the asymptotic optimality of $\tilde{\lambda}_{t}$ in
the case of discrete sampling at the sensors. For this, the sampling
period $h$ must converge to $0$ at an appropriate rate relative to
the communication rate and the horizon of observations, which we
specify in the following theorem.
%
%
\begin{theorem} \label{theo4}
If $t,\Delta^{i}\rightarrow\infty$ and $h \rightarrow0$ so that
\[
\Delta^{i}=o(\sqrt{t}) \quad\mbox{and}\quad \sqrt[3]{h}=o\bigl(
\Delta^{i}/ \sqrt{t}\bigr) \qquad\forall1 \leq i\leq K,
\]
then $\sqrt{A_{t}} (\tilde{\lambda}_{t}-\lambda) \rightarrow\cN(0,1)$,
that is, $\tilde{\lambda}_{t}$ is an asymptotically optimal estimator.
\end{theorem}
\begin{pf}
Since $\sqrt{A_{t}} (\hat{\lambda}_{t}-\lambda) \sim\cN(0,1)$, it
suffices to show that
$\sqrt{A_{t}} |\tilde{\lambda}_{t}-\hat{\lambda}_{t}|$
converges to
$0$ in probability.
Indeed, from (\ref{essence00}) and the fact that $A_{t}$ is a linear
function of $t$,
%
%
\begin{equation}
\label{essence1} \sqrt{A_{t}} \Expel\bigl[|\tilde{\lambda}_{t}-
\hat{\lambda}_{t}|\bigr] \leq\frac{\Theta(\D)}{\sqrt{t}} + \sum
_{i=1}^{K} \frac{\Expel[ \eta_{1}^{i}
]}{\Theta(\Delta^{i}/\sqrt{t})}.
\end{equation}
The first term in the right-hand side goes to 0 if $\D=o(\sqrt{t})$.
The second term goes to $0$ if $\Expel[\eta^{i}_{1}]=o( \Delta^{i}/
\sqrt{t})$ for every $1 \leq i \leq K$. For the latter, it suffices that
$\sqrt[3]{h}=o( \Delta^{i}/ \sqrt{t})$ for every $1 \leq i \leq K$,
due to
(\ref{estim}), which completes the proof.
\end{pf}

\begin{Remark*}
If each $\Delta^{i}$ is fixed as $t \rightarrow \infty$, then Theorem
\ref{theo4} implies that $\hat{\lambda}_{t}$ is asymptotically
efficient as $t \rightarrow\infty$ and $h \rightarrow 0$ so that
$\sqrt[3]{h} \sqrt{t} \rightarrow0$.
\end{Remark*}

\section{Conclusions}\label{sec5}
In this work we considered a parameter estimation problem assuming that
the statistician collects data from dispersed sensors, which
observe continuous (possibly correlated) semimartingales with linear
drifts with respect to a common, unknown parameter.
Motivated by sensor network applications, which are typically
characterized by limited communication bandwidth, we required that the
sensors must send a
small number of bits per transmission and that they should avoid a high
rate of communication with the fusion center.

We proposed a novel methodology for this problem, according to which
the sensors transmit to the fusion center one-bit messages at first
exit times of appropriate statistics that they observe locally. The
fusion center then combines these messages and constructs an estimator
that imitates the optimal centralized estimator (which can be computed
only if there is full access to the sensor observations).

We proved that the resulting estimator is consistent and, for a large
class of processes, asymptotically optimal, in the sense that it
attains the performance of the optimal centralized estimator when a
sufficiently large horizon of observations is available.
However, it is much more efficient from a practical point of view, as
it reduces dramatically the congestion in the network and the
computational burden at the fusion center. This is the case because it
requires the transmission of only one-bit messages from the sensors
and its statistical properties are preserved even with an
asymptotically low rate of communication.

It remains an open problem to design estimators with analogous
optimality properties in more complicated setups, such as when there is
not an explicit form for the optimal centralized estimator, the
dimensionality of the parameter space is large or the sensors take
non-i.i.d., discrete-time observations.

\section*{Acknowledgments}

The author would like to thank Dr. George V. Moustakides and Dr.
Alexandra Chronopoulou for their feedback. Moreover, the author is
grateful to the two anonymous referees and the Associate Editor for
their valuable remarks and suggestions that led to a significant
improvement of earlier versions of this work.



\printaddresses

\end{document}